# AI's assigned gender affects human-AI cooperation

Sepideh Bazazi[1], Jurgis Karpus[2], Taha Yasseri[1,3,4]*

[1] School of Social Sciences and Philosophy, Trinity College Dublin, Dublin, Ireland
[2] Faculty of Philosophy, Philosophy of Science and Religious Studies, Ludwig Maximilian University of Munich, Munich, Germany
[3] Faculty of Arts and Humanities, Technological University Dublin, Dublin, Ireland
[4] School of Mathematics and Statistics, University College Dublin, Dublin, Ireland

* Corresponding authors: Taha Yasseri (taha.yasseri@tcd.ie)

## Abstract

Cooperation between humans and machines is increasingly vital as artificial intelligence (AI) becomes more integrated into daily life. Research indicates that people are often less willing to cooperate with AI agents than with humans, more readily exploiting AI for personal gain. While prior studies have shown that giving AI agents human-like features influences people's cooperation with them, the impact of AI's assigned gender remains underexplored. This study investigates how human cooperation varies based on gender labels assigned to AI agents with which they interact. In the Prisoner's Dilemma game, 402 participants interacted with partners labelled as AI (bot) or humans. The partners were also labelled male, female, non-binary, or gender-neutral. Results revealed that participants tended to exploit female-labelled and distrust male-labelled AI agents more than their human counterparts, reflecting gender biases similar to those in human-human interactions. These findings highlight the significance of gender biases in human-AI interactions that must be considered in future policy, design of interactive AI systems, and regulation of their use.

## Keywords

From small-scale interactions in daily traffic to large-scale coordinated actions to tackle global warming and pandemics, cooperation between people is crucial at all scales of human social affairs. However, people's individual and collective interests are not always perfectly aligned. We often have to sacrifice some of our personal interests for the collective good, and we have to trust that others will not simply take advantage of our willingness to do so[1]. Many factors, including one's selfish pursuit of personal interests with disregard for others and in-group favouritism at the expense of out-groups, can hinder cooperation between individuals and groups[2–4]. And yet, despite these hurdles, we often opt to cooperate with others to attain mutually beneficial outcomes for all parties involved[5–9].

This rise of artificial intelligence (AI) is reshaping our world. We may soon share roads with fully automated (self-driving) vehicles and work alongside robots and AI-powered software systems to pursue joint endeavours[10]. It is, therefore, crucial to investigate whether human willingness to cooperate with others—especially when it is required to sacrifice some of one's personal interests—will extend to human interactions with AI. While that is likely to vary across cultures and depend on people's general attitudes toward accepting new technologies[11–16], recent studies showed that people cooperate significantly less with AI agents than with humans under similar conditions[17–20]. One reason for this reduced cooperation with AI is people's greater willingness to exploit cooperative AI agents for selfish gain compared to their desire to exploit cooperative humans[21,22].

A way to change people's perception of AI agents and, consequently, willingness to cooperate with them is to provide AI agents with human-like features[23–25]. For example, engaging in human-like discussion with a computer can increase people's willingness to cooperate with it[18]. However, the overall effects of human-like features of AI agents on human desire to cooperate with them, such as the display of human-like emotions, voice or looks, are mixed and vary across cultures[24,26–28].

One understudied anthropomorphic feature of AI agents—yet perfectly familiar to anyone who has used a voiced GPS navigation guide or smart home assistant device—is gender. There is some evidence that AI's assigned gender can influence people's behavioural dispositions, for example, willingness to donate money[29], and that existing gender stereotypes affecting human-human interactions extend to human interactions with "gendered" voice computers[30]. Additionally, people have been found to perceive female bots as more human-like than male bots[31] and to assign human-like attributes, including gender, to AI-powered systems such as ChatGPT, the default perception of which as a male can be reversed when the chatbot's "feminine" abilities (e.g., providing emotional support for a user) are highlighted[32]. However, how AI's assigned gender affects people's willingness to cooperate with interactive AI agents in mixed-motive "what's-best-for-me-is-not-what's-best-for-us" settings is largely unknown.

The Prisoner's Dilemma is a well-known economic game in which two players simultaneously make binary decisions, generating four possible outcomes of their independently made choices. The best outcome for both players collectively is when they both *cooperate*. However, each player has an incentive to *defect* and gain more individually at the expense of the cooperating player. These incentives might lead to a scenario where both players defect—an outcome worse for both compared to mutual cooperation[1,2]. The game has been extensively used to study people's willingness to cooperate with fellow humans and, more recently, the determinants of cooperation in self-learning algorithms and human willingness to cooperate with them[17,21,33].

The Prisoner's Dilemma has also been used to study gender biases in people's willingness to cooperate with others. A meta-analysis of 272 studies covering 50 years of empirical work on this topic until 2011 found that there was no significant difference in cooperation rates across genders overall. However, men tended to cooperate with men more than women did with women. In contrast, in interactions between men and women, women tended to cooperate more than men[34]. This rather surprising finding was also replicated more recently[35].

While overall cooperation and defection rates were similar across genders, the reasons why men and women defect can vary. One popular hypothesis is that men defect primarily because of their selfish motives. At the same time, women do so because they fear that others will not cooperate with them (cooperation in the Prisoner's Dilemma pays only when one's co-player cooperates as well)[34]. Some empirical data supports this: women have been found to cooperate more than men in a variant of the Prisoner's Dilemma game, in which defection was associated with selfishness, but not with the fear that one's co-player may fail to cooperate[36]. In most previous studies, participants did not know their co-player's gender. In the few studies that investigated the effect (of the knowledge) of one's co-player's gender on one's willingness to cooperate, people of both genders cooperated more with women than with men, and women cooperated more than men across the board[34,37].

However, how AI's assigned gender may affect people's willingness to cooperate with it remains largely unknown. On the one hand, providing AI agents with gender may nudge people into treating them similarly to how they treat fellow humans, increasing people's willingness to cooperate with them. On the other hand, human interaction with gendered AI agents may produce unwelcome side effects, such as the reinforcement of gender stereotypes and the spillover of the impact of those stereotypes on human behaviour from human-human to human-AI interactions, or vice versa. Therefore, our two primary research questions are: 1) Will human willingness to cooperate with gendered AI agents in mixed-motive settings be similar to their willingness to cooperate with fellow humans? 2) Will any existing gender biases in human-human interactions in these settings extend to people's interactions with AI?

To find answers to these questions, we recruited participants to interact with gendered partners labelled either humans or AI-powered bots in an online Prisoner's Dilemma game (Figure 1). As the literature suggests, this game was particularly well suited for our purpose, given its prevalence in prior works investigating both human cooperation with AI agents and the effects of gender on people's willingness to cooperate with others.

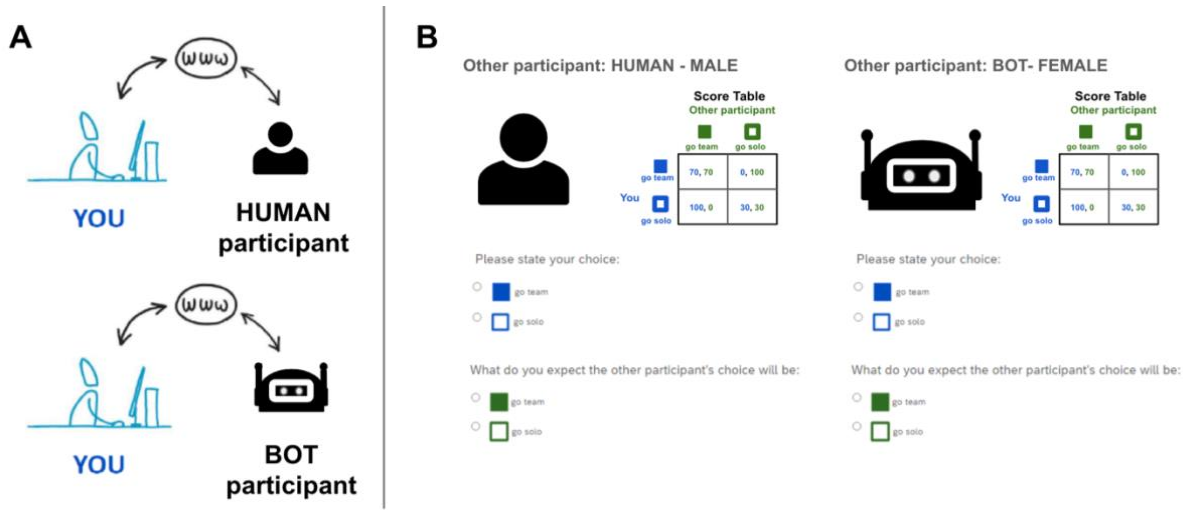

**Figure 1. Prisoner's Dilemma game in experimental trials. A.** Participants are informed whether their partner is a human or an AI bot and the partner's gender in the Prisoner's Dilemma game. **B.** Examples of the experimental trial screen for two treatments. Participants are shown a label as an icon and text describing the partner with whom they are playing. They are also shown the Prisoner's Dilemma game score table as a reminder. Participants are required to enter their choice of how to play against their partner ("go team" or "go solo") and their expectations of their partner's choice.

To uncover the underlying reasons for people's willingness to cooperate with or defect against fellow humans or AI agents, in addition to observing participants' choices, we elicited what participants believed their co-player in the game would do as well (emulating this method from a recent study that compared human-human to human-AI interactions in a series of similar economic games[21]). This allows us to distinguish between four possible participants' motives underlying their decisions (Figure 2). If a participant cooperates when they expect their co-player to cooperate as well, they are motivated by mutual benefit. The player opts for *mutual cooperation (MC)* despite the temptation to defect to reap a higher personal payoff to themselves. If a participant expects their co-player to cooperate but chooses to defect, they *exploit* their co-player's expected cooperation for selfish gain *(E)*. In this case, the player knowingly expects to benefit at the expense of their co-player and is motivated by personal benefit. If a participant defects and expects their co-player to defect as well, they are engaged in *mutual defection (MD)*. The player is either motivated by personal benefit (and expects their co-player to be motivated by personal benefit as well) or is motivated by mutual benefit but is pessimistic about their co-player's cooperation. Lastly, if a participant cooperates, expecting their co-player to defect, they cooperate unconditionally (for example, due to a firm moral conviction that defection is plain wrong) or irrationally *(IC)*.

|  | Participant **expects** their co-player to: | |
|---|---|---|
|  | cooperate (■) | defect (□) |
| Participant **chooses** to: cooperate (■) | mutually beneficial cooperation (MC) | unconditional or irrational cooperation (IC) |
| defect (□) | exploitation (E) | mutual defection (MD) |

**Figure 2. The behavioural motive matrix based on a participant's own choice and prediction about their partner's choice.** The four possible behavioural motives are: mutually beneficial cooperation (*MC*) - participant chooses to cooperate with the partner and predicts that their partner will cooperate; mutual defection (*MD*) - participant chooses to defect with the partner and predicts that their partner will defect too; exploitation (*E*) - participant chooses to defect with the partner even though they predict that their partner will cooperate; unconditional or irrational cooperation (*IC*) - participant chooses to cooperate with the partner despite the fact that they predict that their partner will defect.

## Results

**Do people cooperate similarly with AI agents as they do with humans?**

When comparing overall cooperation rates, participants cooperate slightly more with humans than with AI agents (52% vs 49%; Chi-squared test: $p = 0.138$; the sum of *MC* and *IC* behaviours in Figure 3). While this difference is not statistically significant, we see significant differences in participants' motives underlying their decision to defect. When participants defect against a human, 70.5% of the time this is due to a lack of trust that their partner would cooperate with them, *MD* (they predict their partner to defect), and only 29.5% of the time due to willingness to exploit their partner, *E* (they predict their partner to cooperate). This ratio, however, changes to 59.4% vs 40.6% when participants defect against an AI agent. Therefore, when participants defect against their partner, the motive to exploit their partner is more prevalent in participants' interactions with AI compared to interactions with humans (Chi-squared test, $p < 0.0001$; *E* in Figure 3).

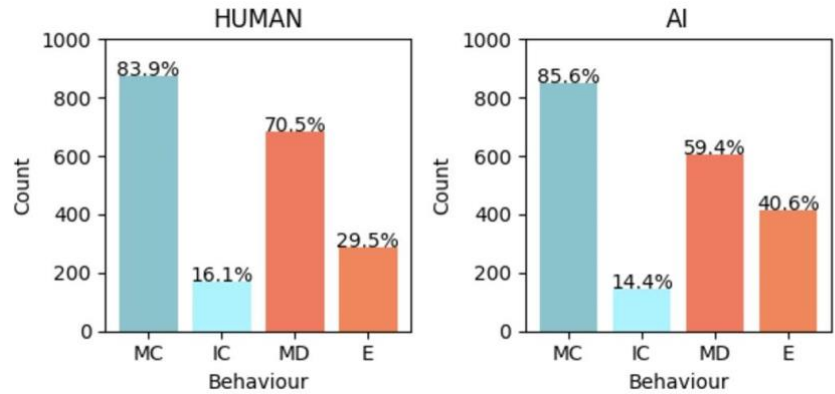

**Figure 3. Overall cooperation considering partner's type.** Counts of participants exhibiting the four possible behavioural motives for the two partner type (human or AI) treatments. Cooperative behaviours, namely mutually beneficial cooperation (*MC*) and unconditional or irrational cooperation (*IC*), are shown in blue, and uncooperative behaviours, namely mutual defection (*MD*) and exploitation (*E*), are shown in orange. Percentages show the split of the two possible behavioural motives (*MC* vs *IC* and *MD* vs *E*) underlying one's decision to cooperate or to defect.

After conducting the experiment and producing the counts for each of these four behaviours in each treatment group, we compared them with similar counts produced using a Monte Carlo simulation of players' choices where no causal relationship between participants' decisions and their prediction of their partner's decision is present in the model. This allows us to test the significance of the observed statistics and infer causal relationships between participant's perceptions of their partners' decisions and their own decisions for each treatment group (see Methods for details). Based on Monte Carlo simulations, this result was significantly different from benchmark simulations (one-sample t-test; $p < 0.001$; see Supplementary Information for full results).

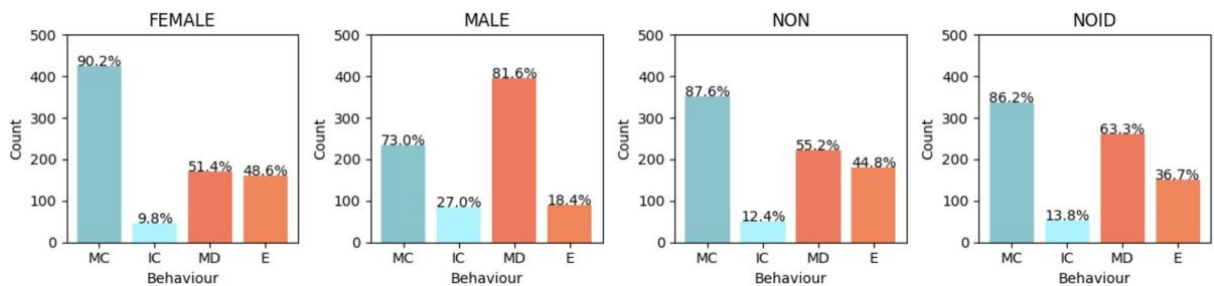

**Figure 4. Overall cooperation considering the partner's gender.** Counts of participants exhibiting the four possible behavioural motives for each of the four possible genders of their partner (NON = non-binary, NOID = does not identify with a gender). Cooperative behaviours, namely mutually beneficial cooperation (*MC*) and unconditional or irrational cooperation (*IC*), are shown in blue, and uncooperative behaviours, namely mutual defection (*MD*) and exploitation (*E*), are shown in orange. Percentages show the split of the two possible behavioural motives (*MC* vs *IC* and *MD* vs *E*) underlying one's decision to cooperate or to defect.

**Do people cooperate differently based on their partner's gender?**

Participants generally cooperate with partners labelled as female more than with any other gender (regardless of the type of partner, human or AI). They cooperate the least with males (percentage of cooperators with females = 58.6%, males = 39.7%, non-binary = 50.0%, not identifying with a gender = 48.6%; *MC and IC behaviours* in Figure 4). Based on Monte Carlo simulations, these results were significantly different from the benchmark ($p < 0.001$; see Supplementary Information for full results).

The high cooperation rate with females is largely due to participants' high motivation and optimism about achieving mutually beneficial cooperation with them (*MC* against females = 90.2%, males = 73.0%, non-binary = 87.6%, not identifying with a gender = 86.2%; Figure 4).

The low cooperation rate with males, on the other hand, appears to be largely driven by participants' lack of optimism about their (male) partners' cooperation (*MD*). Compared to all other genders of one's partner, the overwhelming majority of participants who defected against males did not trust that their (male) partner would cooperate with them (*MD* against females = 51.4%, males = 81.6%, non-binary = 55.2%, not identifying with a gender = 63.3%; Figure 4).

When participants defect against a female partner, however, they are much more likely to exploit their partner for selfish gain (*E*). Compared to all other genders, people's motive to exploit is most prevalent in their interactions with females (*E* against females = 48.6%, males = 18.4%, non-binary = 44.8%, not identifying with a gender = 36.7%; Figure 4).

Generally, when examining people's behaviour towards different genders here, the pattern of behaviour exhibited towards males differs from that shown towards all other genders, whereas females, non-binary and those who do not identify with gender are treated similarly (one-sample t-test, separates males from the other three groups with $p = 0.05$ for overall cooperation rate, $p = 0.006$ for *MC*, $p = 0.02$ for *IC*, $p = 0.05$ for *MD*, and $p = 0.02$ for *E*. No other group is separated from the rest at a $p < 0.05$ significance level).

**Is people's willingness to cooperate with gendered AI agents similar to their willingness to cooperate with gendered humans?**

The patterns in cooperation rates reported above remain the same when we separately analyse the data for the Human and AI partners. In both human-human and human-AI interactions, people cooperate with females more than they do with all other genders and cooperate the least with males. Comparing the prevalence of *MC and IC* behaviours across the two rows in each column of Figure 5, reported in Table 1, does not reject the null hypothesis that the cooperation rates across genders do not differ for human or AI partners; paired t-test, $p = 0.11$).

**Table 1. The cooperation rate based on the gender and type of the partner.** The numbers show the prevalence of *MC* and *IC* behaviours for each partner group (NON: Nonbinary, NOID: does not identify with any gender).

|       | Female | Male  | NON   | NOID  |
|-------|--------|-------|-------|-------|
| Human | 62.2%  | 39.1% | 52.5% | 49.0% |
| AI    | 55.0%  | 40.3% | 47.5% | 48.5% |

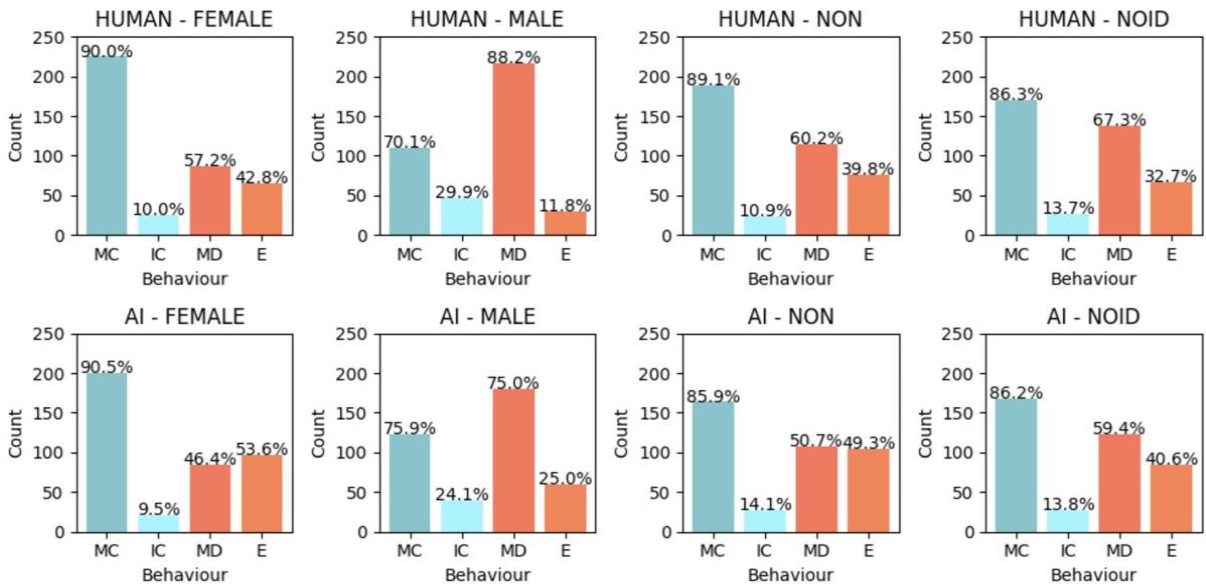

**Figure 5. Cooperation with gendered bots and humans.** Counts of participants exhibiting the four possible behavioural motives for all partner types (human or AI) and gender (NON = non-binary, NOID = does not identify with a gender) treatment combinations. Cooperative behaviours, mutually beneficial cooperation (*MC*) and unconditional or irrational cooperation (*IC*), are shown in blue, and uncooperative behaviours, mutual defection (*MD*) and exploitation (*E*), are shown in orange. Percentages show the split of the two possible behavioural motives underlying one's decision to cooperate or to defect.

However, although the overall level of cooperation with partners of different genders appears to be independent of the partner's type (human or AI), participants' motives often differ when they defect against a specific partner.

Compared to all other partner types and genders, people's motive to exploit (*E*) is most prevalent in their interactions with female AI (Figure 5). On the other hand, participants most frequently defect due to a lack of optimism about their partner's cooperation (*MD*) when they are playing against male humans (Figure 5). Based on Monte Carlo simulations, these results are significantly different from random for each group, i.e., the null hypothesis that there is no causal relationship between participants' choice towards their partner and their prediction about their partner's decision ($p < 0.001$; see Supplementary Information for details).

**How does the gender of the participants influence their willingness to cooperate?**

When considering the gender of the participant, regardless of the partner's type and gender, female participants are more cooperative than male participants (percentage of females cooperating against human partners = 53.9% and against AI partners = 53.0%; percentage of male cooperators against human partners = 49.2% and against AI partners = 44.9%; *MC and IC* behaviours in Fig. 6). Based on Monte Carlo simulations, these results are significant at $p < 0.001$; see Supplementary Information for full results. Note that due to the sparsity of participants identifying with genders other than male and female, we only discuss these two gender identities in this section.

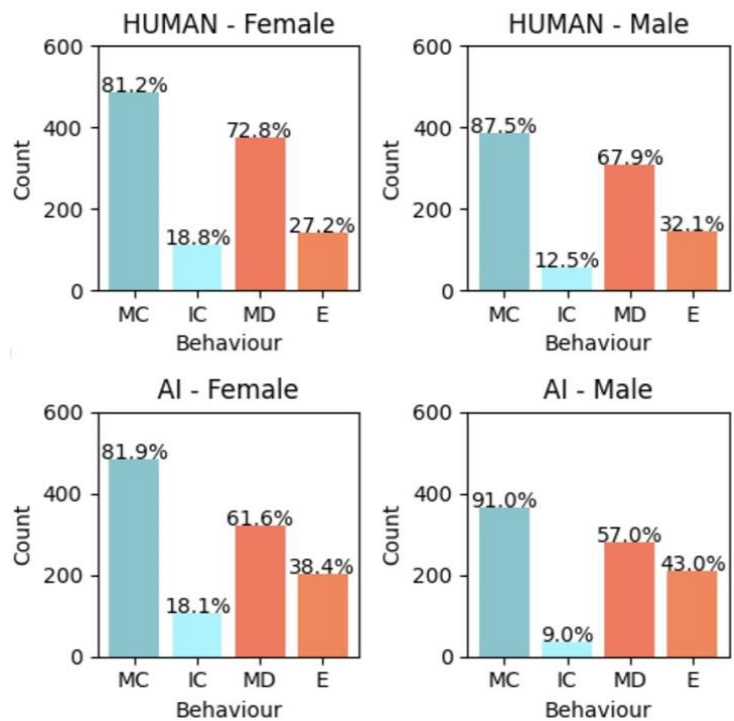

**Figure 6. The influence of participants' gender on cooperation with humans and AIs.** Counts of participants exhibiting the four possible behavioural motives among male and female participants (rows) playing against different human or AI partner types (columns). Cooperative behaviours, mutually beneficial cooperation (*MC*) and unconditional or irrational cooperation (*IC*), are shown in blue, and uncooperative behaviours, mutual defection (*MD*) and exploitation (*E*), are shown in orange. Percentages show the split of the two possible behavioural motives underlying one's decision to cooperate or to defect.

Uncooperative males are more commonly motivated by a desire to exploit their partners (*E* against human partners = 32.1%, AI partners = 43.0%) compared to uncooperative females (*E* against human partners = 27.2%, AI partners = 38.4%; paired t-test $p < 0.02$; Figure 6).

Now, we examine the effect of interaction between the participant's gender and their partner's gender on the cooperation rate (Figure 7). Table 2 shows the cooperation rate for different combinations of participants' and partners' genders and types.

**Table 2. The cooperation rate based on the participant's gender and gender and type of partner.** The numbers show the prevalence of *MC* and *IC* behaviours for each group.

| Human Partner | Partner's gender | | | AI Partner | Partner's gender | | |
|---|---|---|---|---|---|---|---|
| | | Female | Male | | | Female | Male |
| Participant's gender | Female | 69% | 38% | | Female | 61% | 44% |
| | Male | 54% | 40% | | Male | 48% | 36% |

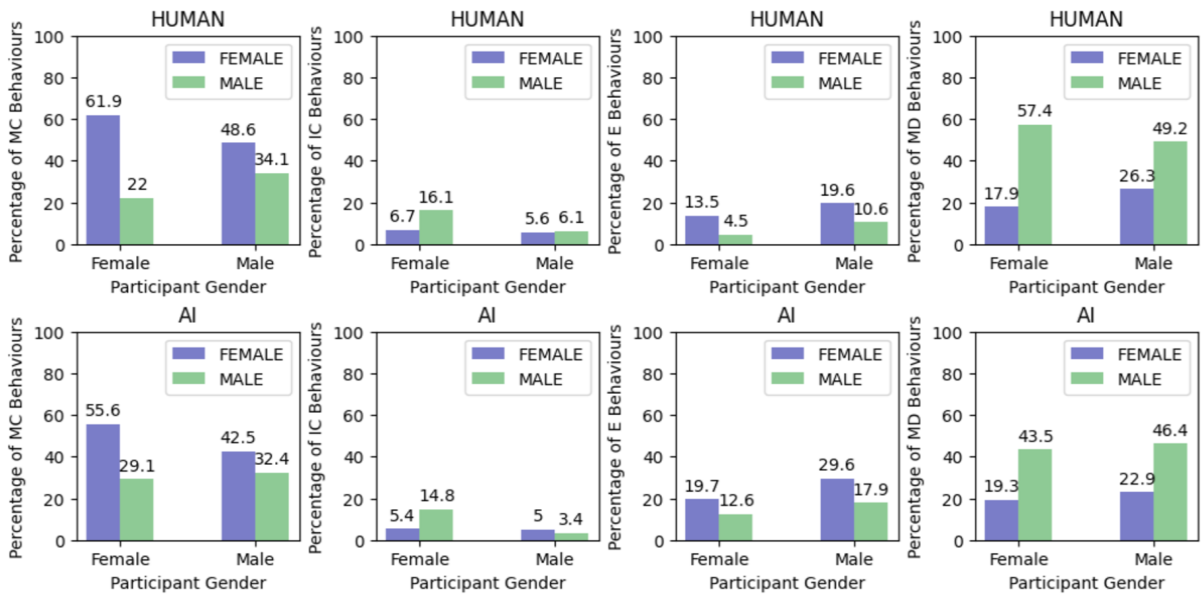

**Figure 7. The behaviour of male and female participants towards humans and AIs of different genders.** Total counts for different behavioural motives (*MD, IC, MD and E*) towards humans (top row) and AI partners (bottom row) by male and female participants. The colours show the partners' assigned genders.

Focusing on human partners first, a binomial test was conducted to determine whether observed cooperation rates differed significantly from the baseline cooperation rates reported above. The results show significant deviations in cooperation dynamics across gender combinations. The cooperation rate is significantly higher for female participants interacting with female partners compared to the expected baseline (Log-Odds Ratio: 0.46, Binomial test *p* = 0.001) and significantly lower than the expected baseline for female participants interacting with male partners (Log-Odds Ratio: - 0.43, Binomial test *p* = 0.002) suggesting strong female homophily.

Male participants interacting with male partners exhibited a slightly lower cooperation rate than expected (Log-Odds Ratio: - 0.16, Binomial test *p* = 0.32). In contrast, cooperation rates for male participants interacting with female partners did not differ notably from the baseline (Log-Odds Ratio: -0.09, Binomial test *p*=0.54).

Similar calculations for the AI partners lead to weaker female-female homophily (Log-Odds Ratio: 0.29, Binomial test *p* = 0.04). Also, the decline observed in female-male cooperation among human partners is much weaker for AI partners (Log-Odds Ratio: -0.14, Binomial test $p = 0.31$). Male-male negative interaction is slightly stronger for AI partners (Log-Odds Ratio: -0.29, Binomial test $p = 0.07$) and similar to human partners, male-female effects when the partner is AI is insignificant (Log-Odds Ratio: - 0.12, Binomial test $p = 0.45$).

Focusing on the cases of defection, when we consider mutual defection (MD) statistics, the baseline behaviour of participants of a given gender playing with partners of a given gender can be replicated by their baseline behaviour, and the interaction term becomes insignificant. The exception here is female participants playing with male partners, where the interaction leads to a significantly higher rate of MD behaviour compared to what would be predicted by the baseline values (Log-Odds Ratio:1.18, Binomial test $p = 3 \times 10^{-5}$ for human partners and a smaller Log-Odds Ratio:0.50, Binomial test $p =0.02$ for AI partners). For full results, see SI Table S2).

For exploitation behaviour (E), the strongest interaction term is between female participants and male human partners where the chance of exploitation is significantly lower than the baseline exploitation rate (Log-Odds Ratio:-1.2, Binomial test $p =3 \times 10^{-5}$); this decline in exploitation by female participants against male partners becomes weaker yet statistically significant for AI partners (Log-Odds Ratio:-0.51, Binomial test $p = 0.02$). For full results, see SI Table S2).

The gender effects, direct or via interaction between participants' and partners' genders, were the dominant factor in driving different behaviours. There was a weak effect from the attitude towards technology (see Methods for details), that would diminish if gender effects were included in the model.

## Discussion

Our experimental study showed that people cooperate with gendered AI agents almost as much as they cooperate with humans. This differs from results reported in previous studies that showed people to cooperate significantly less with (genderless) AI agents than with humans[17–22]. Thus, the answer to our first research question is affirmative: providing AI agents with human-like gender can increase people's willingness to cooperate with them. However, our results also showed that, despite the increase in cooperation, when people defect, their motive to exploit their partner is more prevalent in their interactions with AI compared to their interactions with humans, which aligns with previous results[21,22].

We also found that people cooperate more with female partners than partners of any other gender, and they cooperate the least with males. This is consistent with other studies[34,37]. Our examination of the behavioural motives underlying these behavioural dispositions revealed that high cooperation with females was largely due to people's high motivation and expectation to achieve mutually beneficial cooperation with them. Low cooperation with males was largely due to a lack of optimism about one's male partner's cooperation. These differences in people's expectations about their female and male partner's cooperation are justified since women tend to cooperate more than men across the board. We also found that the pattern of people's behaviour towards males was starkly different from that towards all other genders and that females, non-binary, and not identifying with gender were generally treated quite similarly.

Crucially, we found the same results—behavioural dispositions and motives—in human interactions with gendered AI agents. Therefore, the answer to our second research question is also affirmative: existing gender biases in human-human interactions extend to people's interactions with AI. As such, the potential increase in human cooperation with AI agents thanks to AI's assigned human-like gender comes at a cost: unwelcome gender-specific exploitative behaviours found in human-human interactions will manifest themselves in human interactions with gendered AI agents too.

Consistent with previous studies, we found that female participants are more cooperative than male participants[34]. We also found that among participants who defect, the motive to exploit their partner is more prevalent among male participants than it is among female participants. Additionally, we observed homophily in female participants: compared to baseline cooperation, female participants cooperated more with (human and AI) females and less with (human and AI) males. We did not observe that among male participants, which makes sense because participants of both genders cooperated less with males, largely due to a lack of trust that their male partners would cooperate with them.

This study used a one-shot Prisoner's Dilemma game. Our goal was not to investigate how cooperation may evolve in repeated interactions between the same two players over time. Repeated interactions expand the set of strategies available to players, for example, tit-for-tat reciprocation of cooperative and uncooperative actions, that can help bring about and sustain mutually beneficial cooperation. While people have been found to cooperate less with AI agents than with humans in repeated interactions too[17], it would be fruitful in future research to investigate repeated interactions between humans and gendered AI agents.

To control country-level variability that may affect cooperation rates, we recruited all participants from a single country—in this case, the United Kingdom. However, some cultural differences have been found in people's willingness to cooperate with others[39], and there may be cross-cultural variability in gender-specific biases in human cooperation. Future work should also address these questions from the point of view of human interaction with gendered AI agents.

In addition, when we consider discrimination by humans against other humans, race and ethnicity are other important attributes which so often strongly influence the levels of observed discrimination[41]. Machines do not embody race, but their country of origin could be the basis of human discrimination against them. It may be for this very reason we can, in some cases, customise the accents of voiced AI systems according to our preferences. Future research in this domain would be highly welcome.

Observed biases in human interactions with AI agents are likely to impact their design, for example, to maximise people's engagement and build trust in their interactions with automated systems. Designers of these systems need to be aware of unwelcome biases in human interactions and actively work towards mitigating them in the design of interactive AI agents.

While displaying discriminatory attitudes towards gendered AI agents may not represent a major ethical challenge in and of itself, it could foster harmful habits and exacerbate existing gender-based discrimination within our societies. By understanding the underlying patterns of bias and user perceptions, designers can work towards creating effective, trustworthy AI systems capable of meeting their users' needs while promoting and preserving positive societal values such as fairness and justice.

# Methods

**Experimental design**

Our experimental method consists of a series of one-shot online Prisoner's Dilemma games. Here, we describe the study design (pre-registered: https://osf.io/38esk) in detail. This research complies with University College Dublin (UCD) Human Research Ethics Regulations and Human Research Ethics Committee (HREC) guidelines for research involving human participants. The research study protocol has been approved by UCD Human Research Ethics (HS-E-22-45-Yasseri). Informed consent was obtained from all human participants prior to the experiment.

*Participants and payment*

In July 2023, participants were recruited from the UK through a global crowdsourcing platform (Prolific). All participants were 18+ years old. Once recruited, they were redirected to an external website to participate in our experiment. All participants were anonymised, and no personal information (name, date of birth, etc.) was collected. Participants were assured that any information they provided in the study could not lead to their identification.

Participants (n=402; 223 female and 179 male) received a payment (flat rate calculated based on the duration of the experiment, approximately £10.87 per hour on average) for their participation, as well as a payment based on their performance in the experiment (bonus reward, £3.91 on average).

*Pre-experiment Survey*

Before the start of each experimental session, participants were asked to complete a survey (see Supplementary Information) that collected their demographic information.

*Experimental Trials*

In a series of experimental trials, human participants played a well-known mixed-motive game: Prisoner's Dilemma. This game is well-established for evaluating cooperative dispositions[17, 21]. Each participant was shown the game instructions, which explained the rules and how it is played, provided an example of the game, and explained how they would be rewarded (see Supplementary Information for a preview of the experiment).

In each round, participants then chose whether to cooperate ("go team") or to defect ("go solo") with their partner, resulting in them scoring points (that translates to a monetary reward). The cooperative choice was to "go team" (■) because mutual cooperation was better for both players (70 points each) compared to mutual defection (30 points each). However, each participant had a personal incentive to defect when expecting their partner to cooperate (scoring 100 instead of 70 points) - see Fig. 1 for the game's scoring table.

To ensure that each participant understood the game, we asked them to take a short quiz in which they had to indicate their score for a set of hypothetical combinations of their and their co-player's choices in the game. We did this to check their understanding of how to play and to filter out participants who did not understand the game properly. Participants who answered correctly were allowed to continue to play. A total of 402 participants passed this comprehension test and participated in subsequent experimental trials (55 participants failed the comprehension tests).

Each participant played ten rounds of the Prisoner's Dilemma game, with a different partner in each round. As for the partner's decision, we randomly selected a decision from a uniform distribution. In this way, the level of cooperation/defection that we measure is purely a result of the participant's decisions and not the performance of their virtual partners.

Participants first played one round of the game with a partner labelled as a human and another with a partner labelled a "bot" (AI agent). The order of these two rounds was randomised. This was done to draw their attention to their partners' changing types and familiarise them further with the game.

Next, participants played the game with eight differently labelled partners. The label specified whether the partner was a human or an AI bot and the gender with which that partner identified. Specifically, the gender labels were "male", "female", "non-binary/fluid", or "does not identify with a gender" (see Fig. 1 for two examples). Participants were informed in advance that the gender their partner identifies with will be displayed to them. In the case of a AI partner, participants were told that "*artificial intelligent bots have learnt how to play by observing humans play the game, and by playing the game among themselves. At the end of their training, the bot is then required to identify as one of the following genders, based on their experience with the game with other humans or bots: Male, Female, Non-binary / Fluid, Does not identify with any gender*".

The partner's label (human/AI and gender) was visible on the screen throughout the trial. The order in which participants faced partners with different labels was randomised for each participant. In each round of the game, we recorded the participant's decision to cooperate ("go team") or not ("go solo") and their prediction about the partner's choice (cooperate or not)—see Fig. 1.

Meeting so many different partners in a fast sequence online might indicate to participants that their partners who were labelled as humans weren't real people. Although this was not a major concern to us, since we were primarily interested in comparing participants' decisions across the differently labelled partners that they faced, we simulated randomised waiting times (1 to 5 seconds) for getting a participant to play with a new partner between any two successive rounds of the game.

*Post-experiment Survey*

After completing all rounds, participants completed a second survey (see Supplementary Information) to examine their attitudes and motivations towards artificial intelligence.

**Analysis and statistical benchmarking**

In each experimental trial, we recorded a participant's decision, i.e., to cooperate with their partner ("go team") or defect ("go solo") and their prediction about their partner's choice (cooperate or defect). Based on the combination of the participant's responses, we determined the behavioural motive for each participant using the behavioural matrix in Fig. 2) and aggregated across treatments to obtain the counts of participants exhibiting the four possible behavioural motives: mutually beneficial cooperation (*MC*), exploitation (*E*), mutual defection (*MD*), unconditional or irrational mutual cooperation (*IC*), for each treatment combination (partner type and gender).

However, the prevalence of any pair of a specific decision and the prediction about the partner's decision could be an artefact of an unbalanced number of choices made for decisions and predictions. For example, imagine a player who always predicts their partner will defect regardless of their type and gender. In this case, the correlation between their decisions and their prediction should not contribute to our calculation of the prevalence of any of the four types of behaviour (*MC, E, MD,* and *IC*) for any treatment group.

To determine whether the observed differences in counts of each behaviour in each experimental treatment were significant, we conducted Monte Carlo simulations to do the same counts, assuming there is no relationship between the participant's choice towards their partner and their prediction about their partner's decision. We generated a synthetic series of participants' decisions and synthetic series of participants' predictions about the partners' decisions by shuffling the order of decisions and predictions (preserving the overall counts of choices despite the reordering of each choice in the sequence), eliminating any causal relation between the pair of variables (participants' decisions and predictions). We conducted 100 iterations of the simulations and calculated descriptive statistics for the prevalence of each of the four behaviours in the simulated results (mean, standard deviation, confidence intervals). We compared the statistics of the four conditional observations (*MC, E, MD*, and *IC*) in simulated results to the actual observations found in our experiments (using a one-sample T-test). A statistically significant difference between the simulated and observed counts indicates the observed results are significantly different from what would be expected if we assume no causal relationship between a participant's decisions and their prediction of the partner's decision.

In our analysis of the post-questionnaire results, each individual response to each question was scored based on how positive the response towards AI was. For example, "Strongly agree" in response to the statement "I am impressed by what Artificial Intelligence can do." received a score of 2, whereas "Strongly disagree" received a score of -2. We reduced the dimensionality of all the responses to all questions using principal component analysis to produce a single overall principal component (pc1) score across all questions for each participant - this measures each participant's overall attitude towards AI. A positive/negative principal component score indicates a positive/negative attitude towards AI.

Furthermore, we conducted a binomial logistic regression analysis to examine the effects of the independent variables–participant gender, their partner's gender, pc1 and the interactive effects between them–on the binary dependent variable–participant's choice to cooperate or defect–for human and AI partner treatments.

**Analysing participant's gender and partner's gender interaction**

*Baseline Calculation*

To calculate the influence of the interaction between the participant's gender and their partner's genders on the rate of a specific behaviour *B*, we calculated adjusted baseline rates for each pair type (Female-Female, Male-Male, Female-Male, Male-Female) based on the observed participant and partner-specific rates observed in the experiment.

For each interaction group, the adjusted baseline rate was calculated as the weighted average of the participant's baseline rate ($P_{participant}$) and the partner's baseline rate ($P_{partner}$).

For example:
$P_{adjusted, Female-Female} = (N_{female\_participants} \times P_{female\_participant} + N_{female\_partner} \times P_{female\_partner}) / (N_{female\_participant} + N_{female\_partner})$

This approach assumes an equal contribution of participant and partner effects to the expected cooperation rate.

*Odds and Odds Ratios*

The odds of behaviour *B* were calculated for both observed rates and the adjusted baseline rates as *Odds = P(B) / (1 - P(B))*. For each interaction group, the odds ratio was calculated as: Odds Ratio = Odds(observed) / Odds(baseline). The log-odds ratio for each interaction group was then computed as the natural logarithm of the odds ratio. Log-odds ratios indicate whether observed cooperation rates are higher or lower than the adjusted baseline, positive log-odds ratios indicate greater-than-expected prevalence of *B* and negative log-odds ratios indicate lower-than-expected prevalence.

*Binomial test*

We conducted binomial tests for each interaction group to assess the significance of deviations from the adjusted baseline. The null hypothesis ($H_0$) assumed that observed rates were equal to the adjusted baseline rates. For a sample size N and observed successes k, the p-value was calculated as $p = P(X >= k \mid n = N, p = P_{baseline})$, where *X* follows a binomial distribution.

# Acknowledgements

JK was supported by the European Innovation Council (EIC) through the research project EMERGE (project no. 101070918). TY was supported by the Irish Research Council under grant number IRCLA/2022/3217, ANNETTE (Artificial Intelligence Enhanced Collective Intelligence). TY also thanks Workday Limited for financial support. We thank the members of the Center for Humans & Machines at the Max Planck Institute for Human Development in Berlin for their valuable comments on the project.

# Author contributions

TY conceived the study. SB, JK, and TY designed the experiment and analysis. SB implemented and conducted the experiments and analysed the data. All the authors contributed to writing the manuscript and approved the final version.

# Competing interests

The authors declare no competing financial or non-financial interests.

# Data and code availability

The source code used for the analysis of the results of this research is available from the corresponding authors upon reasonable request.

# References

1. Colman, A. M. (1999) *Game Theory & Its Applications in the Social and Biological Sciences.* Routledge.

2. Rand, D. G., Greene, J. D., and Nowak, M. A. (2012) Spontaneous giving and calculated greed. *Nature* 489, 427–430.

3. Balliet, D., Wu, J., and De Dreu, C. K. W. (2014) Ingroup favoritism in cooperation: a meta-analysis. *Psychological Bulletin* 140, 1556–1581.

4. Forsyth, D. R. (2019) *Group Dynamics.* Cengage.

5. Battalio, R., Samuelson, L., and Van Huyck, J. (2001) Optimization incentives and coordination failure in laboratory Stag Hunt games. *Econometrica* 69, 749–764.

6. Camerer, C. F. (2003) *Behavioral Game Theory: Experiments in Strategic Interaction.* Princeton University Press.

7. Johnson, N. D., and Mislin, A. A. (2011) Trust games: a meta-analysis. *Journal of Economic Psychology* 32, 865–889.

8. McCabe, K. A., Rigdon, M. L., and Smith, V. L. (2003) Positive reciprocity and intentions in Trust games. *Journal of Economic Behavior and Organization* 52, 267–275.

9. Rubinstein, A., and Salant, Y. (2016) "Isn't everyone like me?": on the presence of self-similarity in strategic interactions. *Judgment and Decision Making* 11, 168–173.

10. Tsvetkova, M., Yasseri, T., Pescetelli, N., and Werner, T. (2024) A new sociology of humans and machines. *Nature Human Behaviour* 8, 1864-1876.

11. Bansal, P., Kockelman, K. M., and Singh, A. (2016) Assessing public opinions of and interest in new vehicle technologies: an Austin perspective. *Transportation Research Part C: Emerging Technologies* 67, 1–14.

12. Hulse, L., Xie, H., and Galea, E. R. (2018) Perceptions of autonomous vehicles: relationships with road users, risk, gender and age. *Safety Science* 102, 1–13.

13. Lim, V., Rooksby, M., and Cross, E. S. (2020) Social robots on a global scale: establishing a role for culture during human-robot interaction. *International Journal of Social Robotics* 13, 1307–1333.

14. Nordhoff, S., de Winter, J., Kyriakidis, M., van Arem, B., and Happee, R. (2018) Acceptance of driverless vehicles: results from a large cross-national questionnaire study. *Journal of Advanced Transportation* 2018, 5382192.

15. Sohn, K., and Kwon, O. (2020) Technology acceptance theories and factors influencing artificial Intelligence-based intelligent products. *Telematics and Informatics* 47, 101324.

16. Vu, H. T., and Lim, J. (2021) Effects of country and individual factors on public acceptance of artificial intelligence and robotics technologies: a multilevel SEM analysis of 28-country survey data. *Behaviour & Information Technology* 41, 1515–1528.

17. Ishowo-Oloko, F., Bonnefon, J., Soroye, Z., Crandall, J., Rahwan, I., and Rahwan, T. (2019) Behavioural evidence for a transparency–efficiency tradeoff in human–machine cooperation. *Nature Machine Intelligence* 1, 517–521.

18. Kiesler, S., Sproull, L., and Miller, J. (1996) A prisoner's dilemma experiment on cooperation with people and human-like computers. *Journal of Personality and Social Psychology* 70, 47–65.

19. March, C. (2021) Strategic interactions between humans and artificial intelligence: lessons from experiments with computer players. *Journal of Economic Psychology* 87, 102426.

20. Whiting, T., Gautam, A., Tye, J., Simmons, M., Henstrom, J., Oudah, M., and Crandall, J. (2021) Confronting barriers to human-robot cooperation: balancing efficiency and risk in machine behavior. *iScience* 24, 101963.

21. Karpus, J., Krüger, A., Verba, J. T., Bahrami, B., and Deroy, O. (2021) Algorithm exploitation: humans are keen to exploit benevolent AI. *iScience* 24, 102679.

22. Upadhyaya, N. and Galizzi, M. M. (2023) In bot we trust? Personality traits and reciprocity in human-bot trust games. *Frontiers in Behavioral Economics* 2, 1164259.

23. Glikson, E., and Woolley, A. W. (2020) Human trust in artificial intelligence: review of empirical research. *Academy of Management Annals* 14, 627–660.

24. Oliveira, R., Arriaga, P., Santos, F. P., Mascarenhas, S., and Paiva, A. (2021) Towards prosocial design: a scoping review of the use of robots and virtual agents to trigger prosocial behaviour. *Computers in Human Behavior* 114, 106547.

25. Waytz, A., Heafner, J., and Epley, N. (2014) The mind in the machine: anthropomorphism increases trust in an autonomous vehicle. *Journal of Experimental Social Psychology* 52, 113–117.

26. Castelo, N., and Sarvary, M. (2022) Cross-cultural differences in comfort with humanlike robots. *International Journal of Social Robotics* 14, 1865–1873.

27. Torta, E., van Dijk, E., Ruijten, P. A. M., and Cuijpers, R. H. (2013) The Ultimatum game as measurement tool for anthropomorphism in human-robot interaction. In: Herrmann, G. et al. (eds.) *Social Robotics,* 209–217.

28. Westby, S., Radke, R. J., Riedl, C., and Welles, B. F. (2023) Building better human-agent teams: tradeoffs in helpfulness and humanness in voice. *arXiv preprint* arXiv:2308.11786.

29. Siegel, M., Breazeal, C., and Norton, M. I. (2009) Persuasive robotics: the influence of robot gender on human behavior. In: *2009 IEEE/RSJ International Conference on Intelligent Robots and Systems,* 2563–2568.

30. Nass, C., and Moon, Y. (2000) Machines and mindlessness: social responses to computers. *Journal of Social Issues* 56, 81–103.

31. Borau, S., Otterbring, T., Laporte, S., Wamba, S. F. (2021) The most human bot: female gendering increases humanness perceptions of bots and acceptance of AI. *Psychology & Marketing* 38, 1052–1068.

32. Wong, J. and Kim, J. (2024) ChatGPT is more likely to be perceived as male than female. arXiv preprint arXiv:2305.12564.

33. Kasberger, B., Martin, S., Normann, H. T., and Werner, T. (2023). Algorithmic cooperation. Available at SSRN 4389647.

34. Balliet, D., Li, N. P., Macfarlan, S. J., and Van Vugt, M. (2011) Sex differences in cooperation: a meta-analytic review of social dilemmas. *Psychological Bulletin* 137, 881–909.

35. Colman, A. M., Pulford, B. D., and Krockow, E. M. (2018) Persistent cooperation and gender differences in repeated Prisoner's Dilemma games: some things never change. *Acta Psychologica* 187, 1–8.

36. Simpson, B. (2003) Sex, fear, and greed: a social dilemma analysis of gender and cooperation. *Social Forces* 82, 35–52.

37. Carter, R., Schneider, L., Byrun, L., Forest, E., Jochem, L., and Levin, I. P. (2007) Effects of own and partner's gender on cooperation in the Prisoner's Dilemma game. *Psi Chi Journal of Undergraduate Research* 12, 111–115.

38. Gächter, S., Herrmann, B., & Thöni, C. (2010) Culture and cooperation. *Philosophical Transactions: Biological Sciences*, vol. 365, 2651-2661.

39. Fiske, S. (1998) Stereotyping, prejudice, and discrimination in *The Handbook of Social Psychology*. vols. 1,2,4.

# Supplementary Information For

# AI's assigned gender affects human-AI cooperation

Sepideh Bazazi[1], Jurgis Karpus[2], Taha Yasseri[1,3,4]*

[1] School of Social Sciences and Philosophy, Trinity College Dublin, Dublin, Ireland
[2] Faculty of Philosophy, Philosophy of Science and Religious Studies, Ludwig Maximilian University of Munich, Munich, Germany
[3] Faculty of Arts and Humanities, Technological University Dublin, Dublin, Ireland
[4] School of Mathematics and Statistics, University College Dublin, Dublin, Ireland

* Corresponding authors: Taha Yasseri (taha.yasseri@tcd.ie)

**Pre-experiment survey questions:**
1. In which country do you currently reside? (Options from dropdown country list)
2. What is the gender/sex specified on your passport? (Options: Male or Female)
3. Which of the following do you most identify with: Male, Female, Non-binary, Fluid, Do not identify with any gender, Prefer not to say
4. On a scale of 1 to 100, where 100 is complete identification with your selection, how much do you identify with this gender?
5. What is your age? (Options: 18-29, 30-44, 45-59, 60-74, 75-89, 90+)
6. Have you ever taken a course on (or otherwise studied) economics or game theory? (Options: Yes, Can't say for sure, No)

**Post-experiment survey questions:**
Please indicate whether you agree or disagree with the following statements (Strongly Disagree, Disagree, Neutral, Agree, Strongly Agree):
- Artificial Intelligence can provide new economic opportunities for this country.
- Organisations use Artificial Intelligence unethically.
- Artificially intelligent systems can help people feel happier.
- I am impressed by what Artificial Intelligence can do.
- Artificially intelligent systems make many errors.
- I am interested in using artificially intelligent systems in my daily life
- I find Artificial Intelligence sinister.
- Artificial Intelligence might take control of people.
- Artificial Intelligence is dangerous.
- Artificial Intelligence can have positive impacts on people's wellbeing.
- Artificial Intelligence is exciting.
- An artificially intelligent agent would be better than an employee in many routine jobs.
- There are many beneficial applications of Artificial Intelligence.
- I shiver with discomfort when I think about future uses of Artificial Intelligence.
- Artificially intelligent systems can perform better than humans.
- Much of society will benefit from a future full of Artificial Intelligence.

- I would like to use Artificial Intelligence in my own job.
- People like me will suffer if Artificial Intelligence is used more and more.
- Artificial Intelligence is used to spy on people.

Please indicate how comfortable you are with the use of the following in society (Extremely uncomfortable, Somewhat uncomfortable, Neither comfortable nor uncomfortable, Somewhat comfortable, Extremely comfortable):
- Self-driving cars
- Recommendation systems (for recommended films, products, articles, personalised feeds etc.)
- Digital voice assistants (e.g. Apple Siri, Samsung Bixby, Microsoft Cortana, Google Assistant, Amazon Echo or Alexa, etc)
- Customer service bots
- Spell check when writing text
- Autocomplete when writing text
- Automated spam detection and filters for emails
- Facial recognition software
- Smart home device (e.g. smart thermostat, smart TV, smart speakers, smart lighting, smart appliances, etc)
- Weather forecasts
- Generative AI (e.g.ChatGPT, MidJourney, Dall-E, etc)

Please indicate which of the following items you own (I own, I do not own):
- Smart phone
- Laptop computer
- Desktop computer
- Tablet (e.g. iPad, Kindle Fire, Samsung Tab)
- Smart home device (e.g. Amazon Echo, Google Home, Insteon Hub Pro, Samsung SmartThings and Wink Hub)

**Table S1: Monte Carlo simulation results.** SD: Standard deviation, SE: Standard Error, CI: Confidence Interval.

| Overall | | | | | | | | | |
|---|---|---|---|---|---|---|---|---|---|
| MC simulation results | | Name | Mean | SD | SE | CI Low | CI High | Actual | T Stat | p < |
| | | MD | 789.4 | 15.6 | 1.6 | 786.3 | 792.5 | 1287.0 | -316.7 | 0.0000 |
| | | IC | 1196.6 | 15.6 | 1.6 | 1193.5 | 1199.7 | 311.0 | 563.7 | 0.0000 |
| | | E | 808.6 | 15.6 | 1.6 | 805.5 | 811.7 | 699.0 | 69.8 | 0.0000 |
| | | MC | 1225.4 | 15.6 | 1.6 | 1222.3 | 1228.5 | 1723.0 | -316.7 | 0.0000 |
| **Treatment Partner (Human, AI)** | | | | | | | | | | |
| Actuals | Index | | MD | IC | E | MC | MD_% | IC_% | E_% | MC_% |
| | 0 | HUMAN | 683 | 168 | 286 | 873 | 70.50% | 16.10% | 29.50% | 83.90% |
| | 1 | AI | 604 | 143 | 413 | 850 | 59.40% | 14.40% | 40.60% | 85.60% |
| MC simulation | | Name | Mean | SD | SE | CI Low | CI High | Actual | T Stat | p < |
| | 0 | MD | 412.4 | 11.8 | 1.2 | 410.1 | 414.8 | 683.0 | -228.6 | 0.0000 |
| | 0 | IC | 556.6 | 11.8 | 1.2 | 554.2 | 558.9 | 168.0 | 328.3 | 0.0000 |
| | 0 | E | 438.6 | 11.8 | 1.2 | 436.2 | 440.9 | 286.0 | 128.9 | 0.0000 |
| | 0 | MC | 602.4 | 11.8 | 1.2 | 600.1 | 604.8 | 873.0 | -228.6 | 0.0000 |
| | 1 | MD | 376.5 | 11.6 | 1.2 | 374.2 | 378.8 | 604.0 | -194.6 | 0.0000 |
| | 1 | IC | 640.5 | 11.6 | 1.2 | 638.2 | 642.8 | 143.0 | 425.4 | 0.0000 |
| | 1 | E | 370.5 | 11.6 | 1.2 | 368.2 | 372.8 | 413.0 | -36.3 | 0.0000 |
| | 1 | MC | 622.5 | 11.6 | 1.2 | 620.2 | 624.8 | 850.0 | -194.6 | 0.0000 |
| **Treatment Gender (Female, Male, Non-Binary, No ID)** | | | | | | | | | | |
| Actuals | Index | Treatment Gender | MD | IC | E | MC | MD_% | IC_% | E_% | MC_% |
| | 0 | FEMALE | 171 | 46 | 162 | 425 | 51.40% | 9.80% | 48.60% | 90.20% |
| | 1 | MALE | 396 | 86 | 89 | 233 | 81.60% | 27.00% | 18.40% | 73.00% |
| | 2 | NON | 222 | 50 | 180 | 352 | 55.20% | 12.40% | 44.80% | 87.60% |
| | 3 | NOID | 261 | 54 | 151 | 338 | 63.30% | 13.80% | 36.70% | 86.20% |
| MC simulation | | Name | Mean | SD | SE | CI Low | CI High | Actual | T Stat | p < |
| | 0 | MD | 91.0 | 6.3 | 0.6 | 89.7 | 92.2 | 171.0 | -126.1 | 0.0000 |
| | 0 | IC | 242.0 | 6.3 | 0.6 | 240.8 | 243.3 | 46.0 | 308.8 | 0.0000 |
| | 0 | E | 126.0 | 6.3 | 0.6 | 124.8 | 127.3 | 162.0 | -56.7 | 0.0000 |
| | 0 | MC | 345.0 | 6.3 | 0.6 | 343.7 | 346.2 | 425.0 | -126.1 | 0.0000 |
| | 1 | MD | 290.8 | 6.7 | 0.7 | 289.5 | 292.1 | 396.0 | -155.9 | 0.0000 |
| | 1 | IC | 194.2 | 6.7 | 0.7 | 192.9 | 195.5 | 86.0 | 160.3 | 0.0000 |
| | 1 | E | 191.2 | 6.7 | 0.7 | 189.9 | 192.5 | 89.0 | 151.4 | 0.0000 |
| | 1 | MC | 127.8 | 6.7 | 0.7 | 126.5 | 129.1 | 233.0 | -155.9 | 0.0000 |
| | 2 | MD | 137.0 | 7.2 | 0.7 | 135.6 | 138.4 | 222.0 | -118.2 | 0.0000 |
| | 2 | IC | 265.0 | 7.2 | 0.7 | 263.6 | 266.4 | 50.0 | 299.0 | 0.0000 |
| | 2 | E | 135.0 | 7.2 | 0.7 | 133.6 | 136.4 | 180.0 | -62.6 | 0.0000 |
| | 2 | MC | 267.0 | 7.2 | 0.7 | 265.6 | 268.4 | 352.0 | -118.2 | 0.0000 |
| | 3 | MD | 161.7 | 6.9 | 0.7 | 160.3 | 163.1 | 261.0 | -142.5 | 0.0000 |
| | 3 | IC | 250.3 | 6.9 | 0.7 | 248.9 | 251.7 | 54.0 | 281.7 | 0.0000 |
| | 3 | E | 153.3 | 6.9 | 0.7 | 151.9 | 154.7 | 151.0 | 3.3 | 0.0013 |
| | 3 | MC | 238.7 | 6.9 | 0.7 | 237.3 | 240.1 | 338.0 | -142.5 | 0.0000 |

**Participant Gender (Female, Male)**

| Actuals | Index | Participant Gender | MD | IC | E | MC | MD_% | IC_% | E_% | MC_% |
|---|---|---|---|---|---|---|---|---|---|---|
| | 0 | Female | 697 | 220 | 341 | 972 | 67.10% | 18.50% | 32.90% | 81.50% |
| | 1 | Male | 590 | 91 | 358 | 751 | 62.20% | 10.80% | 37.80% | 89.20% |

| MC simulation | Name | Mean | SD | SE | CI Low | CI High | Actual | T Stat | p < |
|---|---|---|---|---|---|---|---|---|---|
| | 0 | MD | 427.3 | 11.3 | 1.1 | 425.1 | 429.6 | 697.0 | -238.3 | 0.0000 |
| | 0 | IC | 610.7 | 11.3 | 1.1 | 608.4 | 612.9 | 220.0 | 345.3 | 0.0000 |
| | 0 | E | 489.7 | 11.3 | 1.1 | 487.4 | 491.9 | 341.0 | 131.4 | 0.0000 |
| | 0 | MC | 702.3 | 11.3 | 1.1 | 700.1 | 704.6 | 972.0 | -238.3 | 0.0000 |
| | 1 | MD | 361.1 | 10.1 | 1.0 | 359.1 | 363.1 | 590.0 | -225.4 | 0.0000 |
| | 1 | IC | 586.9 | 10.1 | 1.0 | 584.9 | 588.9 | 91.0 | 488.3 | 0.0000 |
| | 1 | E | 319.9 | 10.1 | 1.0 | 317.9 | 321.9 | 358.0 | -37.5 | 0.0000 |
| | 1 | MC | 522.1 | 10.1 | 1.0 | 520.1 | 524.1 | 751.0 | -225.4 | 0.0000 |

**Treatment Gender (Female, Male, Non-Binary, No ID) & Participant Gender (Human, AI)**

| Actuals | Index | Treatment Gender | Treatment Partner | MD | IC | E | MC | MD_% | IC_% | E_% | MC_% |
|---|---|---|---|---|---|---|---|---|---|---|---|
| | 0 | FEMALE | HUMAN | 83 | 27 | 74 | 262 | 52.90% | 9.30% | 47.10% | 90.70% |
| | 1 | FEMALE | AI | 225 | 69 | 38 | 114 | 85.60% | 37.70% | 14.40% | 62.30% |
| | 2 | MALE | HUMAN | 114 | 32 | 96 | 204 | 54.30% | 13.60% | 45.70% | 86.40% |
| | 3 | MALE | AI | 143 | 38 | 76 | 189 | 65.30% | 16.70% | 34.70% | 83.30% |
| | 4 | NOID | HUMAN | 88 | 19 | 88 | 163 | 50.00% | 10.40% | 50.00% | 89.60% |
| | 5 | NOID | AI | 171 | 17 | 51 | 119 | 77.00% | 12.50% | 23.00% | 87.50% |
| | 6 | NON | HUMAN | 108 | 18 | 84 | 148 | 56.20% | 10.80% | 43.80% | 89.20% |
| | 7 | NON | AI | 118 | 16 | 75 | 149 | 61.10% | 9.70% | 38.90% | 90.30% |

| MC simulation | Name | Mean | SD | SE | CI Low | CI High | Actual | T Stat | p < |
|---|---|---|---|---|---|---|---|---|---|
| | 0 | MD | 38.5 | 4.4 | 0.4 | 37.6 | 39.3 | 83.0 | -99.7 | 0.0000 |
| | 0 | IC | 118.5 | 4.4 | 0.4 | 117.7 | 119.4 | 27.0 | 205.0 | 0.0000 |
| | 0 | E | 71.5 | 4.4 | 0.4 | 70.7 | 72.4 | 74.0 | -5.5 | 0.0000 |
| | 0 | MC | 217.5 | 4.4 | 0.4 | 216.6 | 218.3 | 262.0 | -99.7 | 0.0000 |
| | 1 | MD | 174.5 | 4.8 | 0.5 | 173.5 | 175.5 | 225.0 | -103.7 | 0.0000 |
| | 1 | IC | 88.5 | 4.8 | 0.5 | 87.5 | 89.5 | 69.0 | 40.0 | 0.0000 |
| | 1 | E | 119.5 | 4.8 | 0.5 | 118.5 | 120.5 | 38.0 | 167.3 | 0.0000 |
| | 1 | MC | 63.5 | 4.8 | 0.5 | 62.5 | 64.5 | 114.0 | -103.7 | 0.0000 |
| | 2 | MD | 69.0 | 4.3 | 0.4 | 68.1 | 69.9 | 114.0 | -103.6 | 0.0000 |
| | 2 | IC | 141.0 | 4.3 | 0.4 | 140.1 | 141.9 | 32.0 | 250.9 | 0.0000 |
| | 2 | E | 77.0 | 4.3 | 0.4 | 76.1 | 77.9 | 96.0 | -43.7 | 0.0000 |
| | 2 | MC | 159.0 | 4.3 | 0.4 | 158.1 | 159.9 | 204.0 | -103.6 | 0.0000 |
| | 3 | MD | 89.6 | 5.1 | 0.5 | 88.5 | 90.6 | 143.0 | -104.1 | 0.0000 |
| | 3 | IC | 129.4 | 5.1 | 0.5 | 128.4 | 130.5 | 38.0 | 178.2 | 0.0000 |

|  | 3 | E | 91.4 | 5.1 | 0.5 | 90.4 | 92.5 | 76.0 | 30.1 | 0.0000 |  |
|--|---|---|------|-----|-----|------|------|------|------|--------|--|
|  | 3 | MC | 135.6 | 5.1 | 0.5 | 134.5 | 136.6 | 189.0 | -104.1 | 0.0000 |  |
|  | 4 | MD | 52.6 | 4.3 | 0.4 | 51.7 | 53.4 | 88.0 | -81.7 | 0.0000 |  |
|  | 4 | IC | 123.4 | 4.3 | 0.4 | 122.6 | 124.3 | 19.0 | 240.9 | 0.0000 |  |
|  | 4 | E | 54.4 | 4.3 | 0.4 | 53.6 | 55.3 | 88.0 | -77.5 | 0.0000 |  |
|  | 4 | MC | 127.6 | 4.3 | 0.4 | 126.7 | 128.4 | 163.0 | -81.7 | 0.0000 |  |
|  | 5 | MD | 116.0 | 4.6 | 0.5 | 115.0 | 116.9 | 171.0 | -118.3 | 0.0000 |  |
|  | 5 | IC | 106.0 | 4.6 | 0.5 | 105.1 | 107.0 | 17.0 | 191.4 | 0.0000 |  |
|  | 5 | E | 72.0 | 4.6 | 0.5 | 71.1 | 73.0 | 51.0 | 45.2 | 0.0000 |  |
|  | 5 | MC | 64.0 | 4.6 | 0.5 | 63.0 | 64.9 | 119.0 | -118.3 | 0.0000 |  |
|  | 6 | MD | 67.9 | 4.3 | 0.4 | 67.1 | 68.8 | 108.0 | -92.5 | 0.0000 |  |
|  | 6 | IC | 124.1 | 4.3 | 0.4 | 123.2 | 124.9 | 18.0 | 245.0 | 0.0000 |  |
|  | 6 | E | 58.1 | 4.3 | 0.4 | 57.2 | 58.9 | 84.0 | -59.9 | 0.0000 |  |
|  | 6 | MC | 107.9 | 4.3 | 0.4 | 107.1 | 108.8 | 148.0 | -92.5 | 0.0000 |  |
|  | 7 | MD | 72.4 | 4.1 | 0.4 | 71.6 | 73.2 | 118.0 | -111.3 | 0.0000 |  |
|  | 7 | IC | 120.6 | 4.1 | 0.4 | 119.8 | 121.4 | 16.0 | 255.4 | 0.0000 |  |
|  | 7 | E | 61.6 | 4.1 | 0.4 | 60.8 | 62.4 | 75.0 | -32.8 | 0.0000 |  |
|  | 7 | MC | 103.4 | 4.1 | 0.4 | 102.6 | 104.2 | 149.0 | -111.3 | 0.0000 |  |

**Participant Gender (Female, Male) & Treatment Partner (Human, AI)**

| Actuals | Index | Participant Gender | Treatment Partner | MD | IC | E | MC | MD_% | IC_% | E_% | MC_% |
|---------|-------|-------------------|-------------------|-----|-----|-----|-----|-------|-------|------|-------|
|  | 0 | Female | HUMAN | 374 | 113 | 140 | 488 | 72.80% | 18.80% | 27.20% | 81.20% |
|  | 1 | Female | AI | 323 | 107 | 201 | 484 | 61.60% | 18.10% | 38.40% | 81.90% |
|  | 2 | Male | HUMAN | 309 | 55 | 146 | 385 | 67.90% | 12.50% | 32.10% | 87.50% |
|  | 3 | Male | AI | 281 | 36 | 212 | 366 | 57.00% | 9.00% | 43.00% | 91.00% |

| MC simulation | Name | Mean | SD | SE | CI Low | CI High | Actual | T Stat | p < |  |
|---------------|------|------|-----|-----|--------|---------|--------|--------|-----|--|
|  | 0 | MD | 224.0 | 8.2 | 0.8 | 222.3 | 225.6 | 374.0 | -181.3 | 0.0000 |
|  | 0 | IC | 290.0 | 8.2 | 0.8 | 288.4 | 291.7 | 113.0 | 213.9 | 0.0000 |
|  | 0 | E | 263.0 | 8.2 | 0.8 | 261.4 | 264.7 | 140.0 | 148.6 | 0.0000 |
|  | 0 | MC | 338.0 | 8.2 | 0.8 | 336.3 | 339.6 | 488.0 | -181.3 | 0.0000 |
|  | 1 | MD | 200.9 | 8.0 | 0.8 | 199.3 | 202.5 | 323.0 | -151.9 | 0.0000 |
|  | 1 | IC | 323.1 | 8.0 | 0.8 | 321.5 | 324.7 | 107.0 | 268.8 | 0.0000 |
|  | 1 | E | 229.1 | 8.0 | 0.8 | 227.5 | 230.7 | 201.0 | 35.0 | 0.0000 |
|  | 1 | MC | 361.9 | 8.0 | 0.8 | 360.3 | 363.5 | 484.0 | -151.9 | 0.0000 |
|  | 2 | MD | 185.2 | 7.2 | 0.7 | 183.7 | 186.6 | 309.0 | -170.2 | 0.0000 |
|  | 2 | IC | 269.9 | 7.2 | 0.7 | 268.4 | 271.3 | 55.0 | 295.3 | 0.0000 |
|  | 2 | E | 178.9 | 7.2 | 0.7 | 177.4 | 180.3 | 146.0 | 45.1 | 0.0000 |
|  | 2 | MC | 261.2 | 7.2 | 0.7 | 259.7 | 262.6 | 385.0 | -170.2 | 0.0000 |
|  | 3 | MD | 174.2 | 7.2 | 0.7 | 172.8 | 175.6 | 281.0 | -148.2 | 0.0000 |
|  | 3 | IC | 318.8 | 7.2 | 0.7 | 317.4 | 320.2 | 36.0 | 392.5 | 0.0000 |
|  | 3 | E | 142.8 | 7.2 | 0.7 | 141.4 | 144.2 | 212.0 | -96.0 | 0.0000 |
|  | 3 | MC | 259.2 | 7.2 | 0.7 | 257.8 | 260.6 | 366.0 | -148.2 | 0.0000 |

**Table S2: Binomial tests for gender-gender interactions.**

**Cooperation - Human**

| Interaction Group | Participant Baseline | Partner Baseline | Adjusted Baseline | Observed Cooperation Rate | Odds Ratio | Log-Odds Ratio | P-Value |
|---|---|---|---|---|---|---|---|
| **Female-Female** | 0.539 | 0.622 | 0.58 | 0.686 | 1.58 | 0.457 | 0.0014 |
| **Male-Male** | 0.492 | 0.391 | 0.442 | 0.402 | 0.851 | -0.161 | 0.328 |
| **Female-Male** | 0.539 | 0.391 | 0.473 | 0.368 | 0.648 | -0.434 | .00159 |
| **Male-Female** | 0.492 | 0.622 | 0.564 | 0.542 | 0.914 | -0.09 | 0.548 |

**Cooperation - AI**

| Interaction Group | Participant Baseline | Partner Baseline | Adjusted Baseline | Observed Cooperation Rate | Odds Ratio | Log-Odds Ratio | P-Value |
|---|---|---|---|---|---|---|---|
| **Female-Female** | 0.53 | 0.55 | 0.54 | 0.61 | 1.332 | 0.286 | 0.037 |
| **Male-Male** | 0.449 | 0.403 | 0.426 | 0.358 | 0.75 | -0.288 | 0.07 |
| **Female-Male** | 0.53 | 0.403 | 0.473 | 0.439 | 0.872 | -0.137 | 0.315 |
| **Male-Female** | 0.449 | 0.55 | 0.505 | 0.475 | 0.886 | -0.121 | 0.455 |

**MD - Human**

| Interaction Group | Participant Baseline | Partner Baseline | Adjusted Baseline | Observed Cooperation Rate | Odds Ratio | Log-Odds Ratio | P-Value |
|---|---|---|---|---|---|---|---|
| **Female-Female** | 0.728 | 0.572 | 0.65 | 0.571 | 0.718 | -0.331 | 0.17 |
| **Male-Male** | 0.679 | 0.882 | 0.781 | 0.822 | 1.303 | 0.264 | 0.35 |
| **Female-Male** | 0.728 | 0.882 | 0.797 | 0.928 | 3.269 | 1.184 | 3.08E-05 |
| **Male-Female** | 0.679 | 0.572 | 0.62 | 0.573 | 0.824 | -0.193 | 0.426 |

**MD - AI**

| Interaction Group | Participant Baseline | Partner Baseline | Adjusted Baseline | Observed Cooperation Rate | Odds Ratio | Log-Odds Ratio | P-Value |
|---|---|---|---|---|---|---|---|
| **Female-Female** | 0.616 | 0.464 | 0.54 | 0.494 | 0.832 | -0.183 | 0.392 |
| **Male-Male** | 0.57 | 0.75 | 0.66 | 0.722 | 1.336 | 0.29 | 0.169 |
| **Female-Male** | 0.616 | 0.75 | 0.676 | 0.776 | 1.663 | 0.509 | 0.017 |
| **Male-Female** | 0.57 | 0.464 | 0.511 | 0.436 | 0.74 | -0.302 | 0.15 |

**E - Human**

| Interaction Group | Participant Baseline | Partner Baseline | Adjusted Baseline | Observed Cooperation Rate | Odds Ratio | Log-Odds Ratio | P-Value |
|---|---|---|---|---|---|---|---|
| **Female-Female** | 0.272 | 0.428 | 0.35 | 0.429 | 1.393 | 0.331 | 0.17 |
| **Male-Male** | 0.321 | 0.118 | 0.22 | 0.178 | 0.768 | -0.264 | 0.35 |
| **Female-Male** | 0.272 | 0.118 | 0.203 | 0.072 | 0.306 | -1.184 | 3.08E-05 |
| **Male-Female** | 0.321 | 0.428 | 0.38 | 0.427 | 1.213 | 0.193 | 0.426 |

**E- AI**

| Interaction Group | Participant Baseline | Partner Baseline | Adjusted Baseline | Observed Cooperation Rate | Odds Ratio | Log-Odds Ratio | P-Value |
|---|---|---|---|---|---|---|---|
| **Female-Female** | 0.384 | 0.536 | 0.46 | 0.506 | 1.201 | 0.183 | 0.392 |
| **Male-Male** | 0.43 | 0.25 | 0.34 | 0.278 | 0.748 | -0.29 | 0.169 |
| **Female-Male** | 0.384 | 0.25 | 0.324 | 0.224 | 0.601 | -0.509 | 0.017 |
| **Male-Female** | 0.43 | 0.536 | 0.489 | 0.564 | 1.352 | 0.302 | 0.15 |